\documentclass[journal=jpcafh,manuscript=article]{achemso}
\usepackage[utf8]{inputenc}

\usepackage{longtable}
\usepackage{multirow}
\usepackage{amsmath}
\usepackage{subfigure}
\usepackage[usenames,dvipsnames]{xcolor}
\usepackage{soul}
\usepackage{bm}

\usepackage{amssymb}
\usepackage{siunitx}
\usepackage{multicol} \usepackage{wrapfig} \usepackage{enumitem}
\usepackage{bm} \usepackage{outlines} 

\usepackage{url}
\usepackage{hyperref}

\usepackage{siunitx}
\usepackage{booktabs}

\usepackage{xr}

\SectionNumbersOn

\author{Sena Aydin} \affiliation[University of
  Basel]{Department of Chemistry, University of Basel,
  Klingelbergstrasse 80, CH-4056 Basel,
  Switzerland.}
\altaffiliation{These authors contributed equally}

\author{Valerii Andreichev} \affiliation[University of
  Basel]{Department of Chemistry, University of Basel,
  Klingelbergstrasse 80, CH-4056 Basel,
  Switzerland.}
\altaffiliation{These authors contributed equally}

\author{Pantelis Maragkoudakis} \affiliation[University of Basel]{Department of
  Chemistry, University of Basel, Klingelbergstrasse 80, CH-4056
  Basel, Switzerland.}

\author{Markus Meuwly} \affiliation[University of Basel]{Department of
  Chemistry, University of Basel, Klingelbergstrasse 80, CH-4056
  Basel, Switzerland.}  \email{m.meuwly@unibas.ch}

\title{Tripeptide-Dynamics from Empirical and Machine-Learned Energy
  Functions}

\begin{document}

\begin{abstract}
Molecular dynamics simulations for tripeptides in the gas phase and in
solution using empirical and machine-learned energy functions are
presented. For cationic AAA a machine-learned potential energy surface
(ML-PES) trained on MP2 reference data yields quantitative agreement
with measured splittings of the amide-I vibrations. Experimental
spectroscopy in solution reports a splitting of 25 cm$^{-1}$ which
compares with 20 cm$^{-1}$ from ML/MM-MD simulations of AAA in
explicit solvent. For the AMA tripeptide a ML-PES describing both, the
zwitterionic and neutral form is trained and used to map out the
accessible conformational space. Due to cyclization and H-bonding
between the termini in neutral AMA the NH- and OH-stretch spectra are
strongly red-shifted below 3000 cm$^{-1}$. The present work
demonstrates that meaningful MD simulations on the nanosecond time
scale are feasible and provides insight into experiments.
\end{abstract}

\section*{Introduction}
Small peptides are valuable proxies for characterizing and
understanding the structure, dynamics, spectroscopy and thermodynamics
of larger proteins.\cite{stenner:2023} Historically, early nuclear
magnetic resonance (NMR) experiments indicated that short linear
peptides in water exhibit predominantly random distributions of
conformations.\cite{anfinsen:1969} These studies were carried out on
digested fragments of different length of staphylococcal
nuclease. However, subsequent 2-dimensional NMR studies provided
evidence that even the conformational space of tripeptides is
restricted and leads to sampling of well-defined
structures.\cite{wright:1988} Evidence for turn formation in water had
been provided by experiments and simulations on terminally blocked NPY
and YPN tripeptides.\cite{oka:1984} The main interest in these earlier
studies concerned elucidation of protein folding pathways which made
contact in particular with the diffusion model for protein folding by
Weaver and Karplus\cite{weaver:1976} which was later used in more
coarse-grained simulations through solving the Smoluchowski equation
on a precalculated protein folding free energy
landscape.\cite{MM.folding:2007} Finally, short peptides have been
used to establish aromatic-aromatic interactions for protein
stabilization.\cite{petsko:1985}\\

\noindent
One of the most thoroughly researched tripeptides is trialanine
(AAA).\cite{woutersen:2000,Woutersen,Hamm.pnas.2001,Schweitzer-Stenner.jacs.2001,Woutersen:2001,Mu.jpcb.2002,Graf2007,Gorbunov2007,Oh2010,Xiao2014,Tokmakoff2018,MM.ala:2021}
A combined experimental and molecular dynamics (MD) study using
non-linear time-resolved spectroscopy on AAA found conformational
heterogeneity of the peptide.\cite{Woutersen} AAA conformational
ensembles were also studied using two-dimensional IR and NMR
spectroscopies.\cite{Tokmakoff2018,Xiao2014,Oh2010} Two-dimensional IR
studies probed the subpicosecond dynamics\cite{Hamm.pnas.2001} and
with isotopically labelled AAA the dipole-dipole coupling strength was
determined.\cite{Woutersen:2001} A more recent MD study using a
multipolar energy function determined the infrared (IR) spectroscopy
and conformational landscape.\cite{MM.ala:2021} Notably, the dihedral
distributions found from these simulations were consistent with
$(\Phi,\Psi)$ maps based on a Bayesian refinement on the measured and
computed 1d-IR spectra.\cite{Tokmakoff2018} Importantly, Bayesian
refinement does not yield an improved underlying energy function
suitable for molecular simulations but rather provides information on
which parts of the Potential Energy Surface (PES) are probed by the
experiment and require refinement. Such a mapping between
observable(s) (here IR spectrum) and the underlying sampling
(trajectory or wavefunction) is also capitalized on in PES-morphing
approaches which use coordinate scaling techniques to reshape the PES
constrained by measurements.\cite{MM.morph:1999,MM.morph:2024}\\

\noindent
Short peptides have also served as proxies to develop, test, and
refine new experimental and computational techniques. One example is
the alanine-dipeptide which was used as a topical system to identify
reaction coordinates,\cite{dinner:2005} to develop new free energy
techniques,\cite{laio:2002} or to test mixed quantum mechanical
molecular mechanics methods,\cite{field:1990} to name a
few. Tripeptides are a the shortest peptides that can form rudimentary
secondary protein structure motifs. As such they are meaningful
proxies to investigate the dynamics for intramolecular H-bond
formation and for characterizing the hydration (dynamics) around
elongated and compact peptide structures. As such it is of interest to
thoroughly investigate the conformational landscape of short peptide
sequences using the most advanced computational techniques. This is
the purpose of the present work.\\

\noindent
Empirical energy functions are particularly successful models to
investigate a wide range of biological and chemical
systems. Importantly, they provide a meaningful zeroth-order
approximation to the energetics and dynamics of systems spanning a
wide range of spatial and temporal scales. For increased realism and
improved performance it is meaningful to augment empirical energy
functions with additional functionality. This can, e.g., be
accomplished by replacing point charge electrostatics with higher
order atom-centered multipolar electrostatics or by using distributed
charge models. For bonded interactions, Morse oscillators or models
based on reproducing kernels can be employed instead of harmonic
energy functions. Due to the rapid progress in machine learning
(ML)-based techniques it is also conceivable that all bonded
interactions are represented as a neural network (NN). This is the
approach pursued in the present work.\\

\noindent
The present work is structured as follows. First, the methods are
introduced. This is followed by results for hydrated AAA and AMA in
the gas phase using empirical and ML-PESs. For both systems somewhat
different computational approaches are followed to highlight
advantages and shortcomings for designing ML-PESs for systems beyond
individual molecules. Finally, the results are discussed in a broader
context and conclusions are drawn.\\

\section*{Methods}

\subsection*{The Potential Energy Surfaces}
The present work employs two different representations of the
potential energy surface. The first is the standard
CGenFF\cite{cgenff} empirical energy function which was parametrized
together with the TIP3P water model.\cite{tip3p} Secondly, machine
learning-based energy functions were trained based on electronic
structure calculations using a PhysNet neural network representation,
which is described in the following.\\

\noindent
The ML-PES for cationic AAA was generated by sampling from molecular
dynamics (MD) simulations. These simulations were performed in both
the gas and solution phases using CHARMM and CGenFF at 300 K and 500
K.\cite{charmm:2009,MM.charmm:2024} For one part of the simulations in
gas phase, all bonds involving hydrogen atoms were described by a soft
Morse oscillator to more broadly sample these bonds. The resulting
ML-PES will be more robust because at ambient temperatures it is
unlikely that the dynamics will explore out-of-sample structures for
these coordinates. From these simulations a total of 12500 structures
were generated and energies, forces, and dipole moments were
determined at the MP2/6-31G(d,p) level of theory using the MOLPRO
suite of codes.\cite{molpro.2011}\\

\noindent
For the AMA tripeptide, gas-phase MD simulations at 300 K were
performed using CHARMM\cite{charmm:2009,MM.charmm:2024} and the
CGenFF\cite{cgenff} energy function to generate structures for the
zwitterionic form. To generate a diverse set of conformational
samples, Replica Exchange Molecular Dynamics (REMD)
simulations\cite{sugita.cpl:1999} were carried out across a range of
temperatures: 300, 350, 400, 450, 500, 550, 600, and 650
K. Additionally, Morse oscillators were used for all bonds involving
hydrogen atoms and the CO bonds to provide to provide broader sampling
which renders the ML-PES more robust. In total, 20000 samples were
extracted from the REMD simulations for which reference energies,
forces, and dipole moments were determined at the RI-MP2/[def2-SVP +
  def2-SVP/C] level of
theory,\cite{bernholdt.csl:1996,weigend.pccp:2005} using the ORCA
Software.\cite{neese.cms:2025} \\

\noindent
For both tripeptides the reference energies and forces together with
the molecular dipole moments were used to train PhysNet by minimizing
the loss function
\begin{equation}
\label{eq:loss}
\begin{aligned}
\mathcal{L} &= w_E \left| E - E^{\text{ref}} \right| + \frac{w_F}{3N}
\sum_{i=1}^{N} \sum_{\alpha=1}^{3} \left| -\frac{\partial E}{\partial
  r_{i,\alpha}} - F^{\text{ref}}_{i,\alpha} \right| \\ &+ w_Q \left|
\sum_{i=1}^{N} q_i - Q^{\text{ref}} \right| + \frac{w_p}{3}
\sum_{\alpha=1}^{3} \left| \sum_{i=1}^{N} q_i r_{i,\alpha} -
p^{\text{ref}}_{\alpha} \right| + \mathcal{L}_{\text{nh}}.
\end{aligned}
\end{equation}
using the Adam optimizer \cite{kingma:2014,reddi:2019}. The
hyperparameters \cite{MM.physnet:2019,MM.physnet:2023} $w_i$ $i \in \{
E, F, Q, p \}$ differentially weigh the contributions to the loss
function and were $w_E = 1$ [1/energy], $w_F \sim 52.92$
[length/energy], $w_Q \sim 14.39$ [1/charge] and $w_p \sim 27.21$
[1/charge/length], respectively, and the term $\mathcal{L}_{\rm nh}$
is a ``nonhierarchical penalty'' that regularizes the loss function
\cite{MM.physnet:2019}. For training, a 80/10/10~\% split of the data
as training/validation/test sets was used. For the AAA tripeptide a
TensorFlow-based version of PhysNet\cite{MM.physnet:2019} was used for
the optimization whereas for AMA the Asparagus Software was
employed\cite{MM.asparagus:2025}. \\

\subsection*{MD simulations}
Two separate types of simulations were run. For cationic AAA MD
simulations for the hydrated system using CGenFF and the ML-PES were
used whereas for AMA simulations were run in the gas phase. The reason
for this was the fact that AAA has been characterized extensively in
the past with possibilities to compare directly with measurements
whereas for AMA data for comparison is scarce. Molecular dynamics
simulations were run using the CHARMM\cite{charmm:2009} and
pyCHARMM\cite{buckner.jctc:2023} codes employing the
CGenFF\cite{cgenff} and PhysNet/MM energy functions, respectively. The
pyCHARMM code is the python implementation of
CHARMM.\cite{buckner.jctc:2023}\\

\noindent
For the simulations of cationic AAA in solution, the peptide was
solvated in a $41 \times 41 \times 41$ \AA\/$^3$ box of TIP3P
water.\cite{tip3p} The systems were minimized, heated and equilibrated
in the $NpT$ ensemble, followed by production simulations 1.5 ns in
length. A Nos\'{e}-Hoover thermostat and piston (Langevin piston)
together using the Leapfrog algorithm. For the nonbonded
contributions, a distance-based cutoff at 14 \AA\/ was used. For the
ML/MM-MD simulations, the energies and forces were those of the
trained PhysNet models and mechanical embedding was used in the
simulations with water. The Lennard Jones parameters on the atoms
treated with PhysNet were those from CGenFF.\cite{cgenff}\\

\noindent
For the gas phase simulations of AMA the structures were first
optimised using the CGenFF and ML-PES energy functions. Next, random
momenta were drawn from a Maxwell-Boltzmann distribution corresponding
to $T = 300$~K, which were assigned to the atoms. All MD simulations
were carried out in the $NVE$ ensemble using the Velocity Verlet
algorithm\cite{verlet:1967} and a time step of $\Delta t = 1$~fs. as
all bonds, including those involving hydrogen atoms were flexible. The
systems were equilibrated for 100~ps, followed by production
simulations of 200~ps simulation time with saving interval 5~fs.\\

\subsection*{Analysis}
To characterize the accessible conformational space for both
tripeptides, MD simulations with initially constrained $\Phi$ and
$\Psi$ angles were carried out. For this, angles $[\Phi_1,\Psi_1] \in
[-180,180]^\circ$ and $[\Phi_2,\Psi_2] \in [-180,180]^\circ$ were
constrained in intervals of $10^\circ$ during the first 100 ps of the
simulation. After this equilibration period, the constraints were
removed and the system was allowed to evolve freely for 1 ns. Then,
from the unconstrained portion of the trajectory at 1 ns, the
probability distribution $P(\Phi,\Psi)$ was estimated using kernel
density estimation (KDE) \cite{parzen.1962}.\\

\noindent
Radial distribution functions (RDFs or $g(r)$) were determined. The
radial distribution function was computed using VMD \cite{VMD}, based
on trajectory frames saved at 2.5 ps intervals. These saved
configurations were utilized for the statistical evaluation of the
RDF. The cutoff for RDFs is determined as 10.0 \AA\/.\\

\noindent
Trajectory frames saved every 5 fs were used for subsequent
analysis. The fluctuating charges were computed for 6000 sampled
structures, taken every 2.5 ps, and the median values were used for
further evaluation. The calculated fluctuating PhysNet charges were
employed to calculate the IR spectrum from the time-dependent dipole
moment. IR spectra $I(\omega)$ were calculated from the Fourier
transform of the dipole-dipole auto-correlation
function\cite{gordon:1968,berne:2000} according to
\begin{equation}
  I(\omega) \propto Q(\omega) \cdot \mathrm{Im}\int_0^\infty dt\,
  e^{i\omega t} \sum_{i=x,y,z} \left \langle \boldsymbol{\mu}_{i}(t)
  \cdot {\boldsymbol{\mu}_{i}}(0) \right \rangle
\label{eq:IR}
\end{equation}
 Here, $\boldsymbol{\mu}_{i}(t)$ is the molecular dipole moment
  along direction $i$ at time t and $Q(\omega)$ is a quantum
  correction factor \cite{qcorr_2004}
\begin{equation}
Q(\omega) = \tanh\left(\frac{\beta \hbar \omega}{2}\right)        
\end{equation}
\\

\section*{Results and Discussion}
This section presents the results and discusses them vis-a-vis
experiments and earlier simulations. For AAA the conformational
landscape and IR spectroscopy were extensively
investigated\cite{woutersen:2000,Woutersen,Hamm.pnas.2001,Schweitzer-Stenner.jacs.2001,Woutersen:2001,Mu.jpcb.2002,Graf2007,Gorbunov2007,Oh2010,Xiao2014,Tokmakoff2018,MM.ala:2021}
whereas for AMA only a vibrational circular dichroism spectrum was
reported in the past.\cite{eker:2004}\\

\subsection*{The AAA Tripeptide}
For cationic AAA first structural aspects are discussed, followed by
an overall characterization of the underlying folding energy landscape
and the vibrational spectroscopy specifically in the amide-I
region. Two energy functions were used for this: they were the CGenFF
empirical energy function and a neural network-based model using the
PhysNet architecture.\cite{MM.physnet:2019} The performance of this
model is shown in Figure \ref{sifig:aaa-nn}. The RMSE$(E)$ was 0.34
kcal/mol (MAE: 0.41 kcal/mol) on the test set and for the forces the
RMSE$(F)$ was 0.82 (kcal/mol)/\AA\/ (MAE:0.41 (kcal/mol)/\AA\/). All
simulations discussed in this subsection were carried out in explicit
solvent. For the ML/MD simulations, the PhysNet model was used to
describe the ML part, while for the empirical part CGenFF was
employed.\\

\begin{figure}[h!]
    \centering
    \includegraphics[width=1\linewidth]{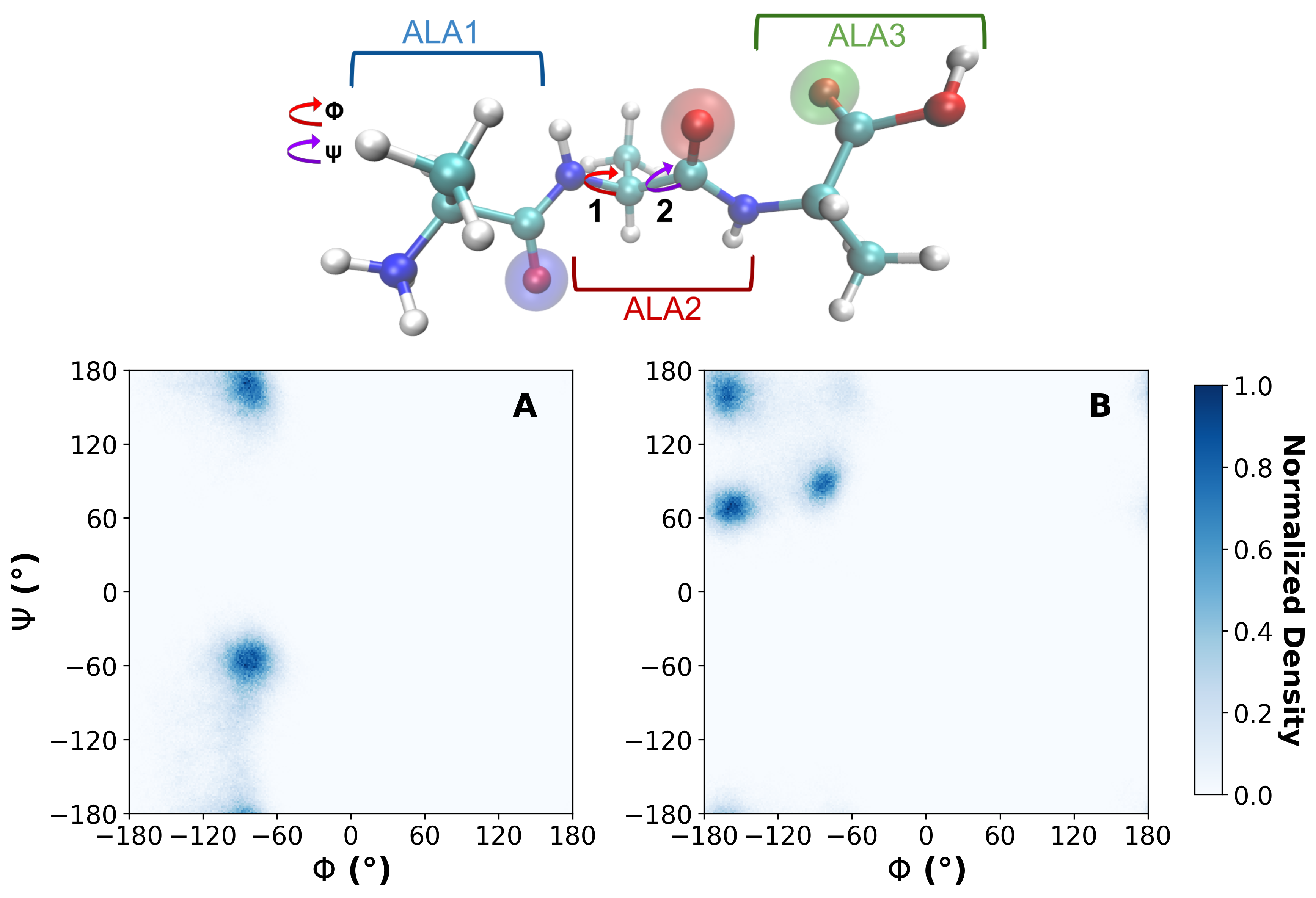}
    \caption{Top: The structure of cationic AAA with ALA1, ALA2, and
      ALA3 labelled. The $\Phi$ (red, 1) and $\Psi$ (lila, 2) angles
      are indicated and the three -CO groups for which the IR spectra
      were determined are highlighted (blue, red, green). Panel A:
      $[\Phi,\Psi]-$angle (Ramachandran) plot for AAA simulations
      using CGenFF. Panel B: $[\Phi,\Psi]-$angles from simulations
      using the PhysNet model in ML/MM-MD simulations. The total
      simulation time was 1.5 ns and the histogram was generated from
      $250 \ 250$ bins.}
    \label{fig:ramachandran_aaa}
\end{figure}

\noindent
Figure \ref{fig:ramachandran_aaa} reports Ramachandran plots for AAA
from simulations using the CGenFF (panel A) and ML/MM (panel B) energy
functions. As AAA consists of two peptide units, the
structure-relevant dihedral angles are $\Phi_1$ and $\Psi_2$ which are
referred to as $[\Phi, \Psi]$ in the following, see top of Figure
\ref{fig:ramachandran_aaa}. As a guide, standard $[\Phi, \Psi]-$values
from protein Ramachandran maps for the $\beta$, PPII, $\alpha_R$, and
$\alpha_L$ conformations are centered at [$-140^\circ$, $130^\circ$],
[$-75^\circ$, $150^\circ$], [$-70^\circ$, $-50^\circ$], and
[$50^\circ$, $50^\circ$], respectively.\\

\noindent
Figure \ref{fig:ramachandran_aaa}A shows that simulations using the
CGenFF empirical energy function primarily populate the PPII
($[-85^\circ, 170^\circ]$) and $\alpha_R$ regions which is consistent
with previous simulations using the same energy
function.\cite{Tokmakoff2018,MM.ala:2021} The population of $\alpha_R$
is also in line with the observation that empirical energy functions
for protein simulations tend to favor helical
conformations.\cite{best:2010} Contrary to that, the region $\Phi \in
[-180,-60]^\circ$ and $\Psi \in [60, 180]^\circ$, characteristic of
$\beta$ and II conformations is populated during the ML/MM-MD
simulations. The distributions become denser toward the $\beta$-sheet
region, with a notable population near [$-160^\circ$, $165^\circ$]. In
these simulations the atom-centered partial charges fluctuate as a
function of geometry which is a design-feature of
PhysNet.\cite{MM.physnet:2019}\\

\begin{figure}[h!]
    \centering
    \includegraphics[width=0.6\linewidth]{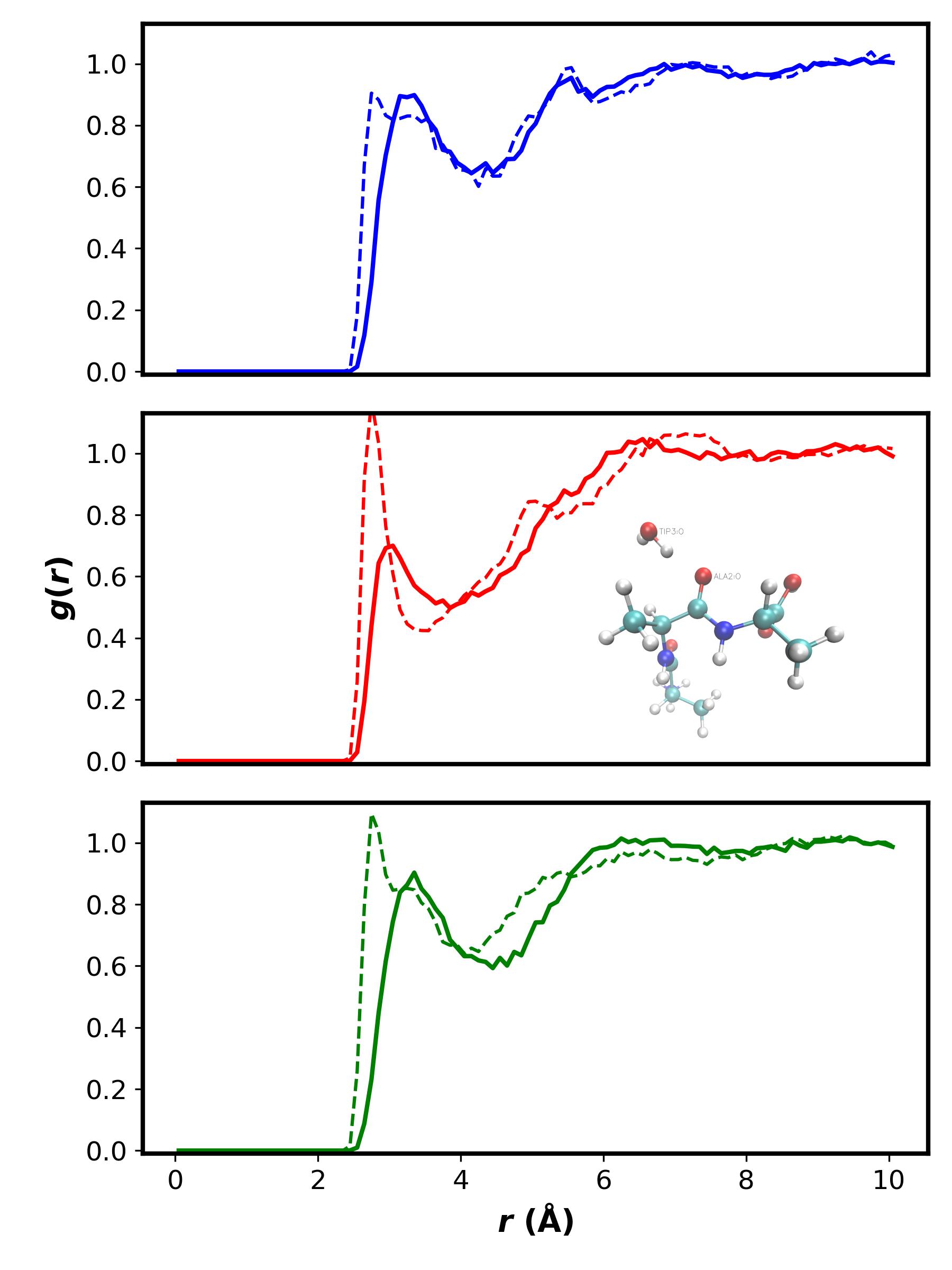}
    \caption{The radial distribution functions, $g(r)$, as a function
      of the distance between the water oxygen atoms, O$_{\rm W}$, and
      O$_{\rm CO}$ atoms for ALA1, ALA2, and ALA3, from top to bottom,
      see Figure \ref{fig:ramachandran_aaa}. Results from simulations
      using CGenFF and the ML/MM energy functions are shown as dashed
      and solid lines. The line colors correspond to the respective
      selected oxygen atoms, see Figure \ref{fig:ramachandran_aaa}.}
    \label{fig:gr}
\end{figure}

\noindent
Earlier work based on Bayesian refinement, guided by experimental
IR-spectroscopy, determined the changes required in the conformational
ensemble characterised by the underlying Ramachandran
map.\cite{Tokmakoff2018} Starting from simulations using the CGenFF
energy function it was found that $\alpha-$helical motifs need to be
removed with concomitant population of the $\beta$ and PPII
conformations in order to improve the match between measured and
calculated IR-spectra. It should, however, be noted that such a
refined Ramachandran map can not be used for MD simulations as this
does not constitute a new energy function. Rather, a Bayesian approach
reweights the underlying population to minimize the difference between
the target (experimental) and computed IR
spectrum.\cite{Tokmakoff2018}\\

\noindent
Another approach was followed in more recent simulation work which
aimed at rational improvements of the underlying energy
function.\cite{MM.ala:2021} Two essential modifications were included:
atom-centered partial charges on cationic AAA were replaced by a
multipolar representation, and the CO-bonds were described as Morse
oscillators instead of harmonic potentials. MD simulations of AAA in
solution using such an improved energy function confirmed that
$\alpha-$helical structures are only populated at the 1-\% level
whereas the $\beta$ and PPII conformations are the dominant regions
sampled. As a comparison, semiempirical-DFT MD simulations confirmed
the absence of helical structures whereas $\beta$ and PPII structures
are primarily sampled.\\

\noindent
Another structural feature of peptides in solution is the exposure of
particular motifs to water. One degree of freedom that is relevant for
AAA in solvent is the water structuring around the backbone --CO
units. Figure \ref{fig:gr} shows the radial distribution function,
$g(r)$, between water oxygen atoms O$_{\rm W}$ and the selected
carbonyl (C=O) oxygen atoms O$_{\rm CO}$ of the tripeptide. The colors
of the carbonyl oxygens in Figure \ref{fig:ramachandran_aaa}
correspond to the data colors in the plots. Results from using the
CGenFF and ML/MM energy functions are shown as dashed and solid lines,
respectively.\\

\begin{figure}[h!]
    \centering
    \includegraphics[width=0.7\linewidth]{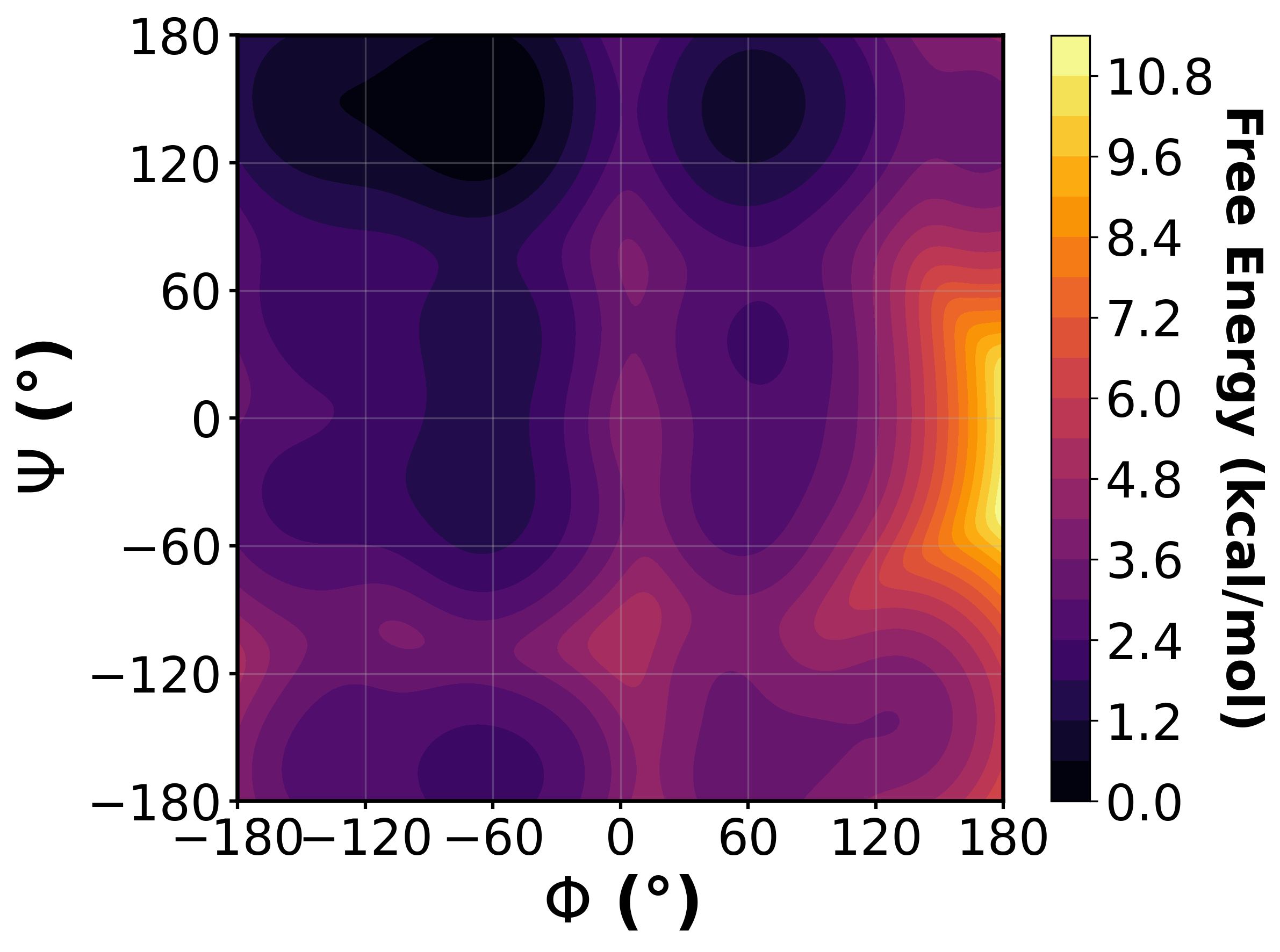}
    \caption{Relaxed pseudo-free energy surface for cationic AAA in
      terms of $\Phi$ and $\Psi$ dihedrals. The underlying probability
      distribution $P(\Phi,\Psi)$ was calculated after relaxation of
      the angles by sampling structures at 1 ns from simulations using
      the CGenFF energy function. This is not an equilibrium free
      energy surface but rather informs about possible low-energy
      regions sampled on the 1 ns time scale.}
    \label{fig:aaa_fes}
\end{figure}

\noindent
For ALA1, the first peak of $g(r)$ from CGenFF peaks at 2.75 \AA\/,
compared with 3.15 \AA\/ from using the ML/MM-PES with equal peak
heights. For larger values of $r$ the structuring of the water is
virtually indistinguishable. Differences are larger for ALA2. Here,
the peak height is considerably larger from simulations using the
empirical energy function compared with simulations using the
ML/MM-PES. Again, the position of the first peak is at larger
separations when using the NN-based PES for the tripeptide (2.75
\AA\/ vs. 3.00 \AA\/). Also, the depth and location of the first
minimum differ. For ALA3, peak heights and location of the first
maximum follow the trends for ALA2 but with less pronounced
differences in the first peak height. The first maximum is at 2.76
\AA\/ compared with 3.34 \AA\/ and the position of the first minimum
differs by 0.5 \AA\/ whereas its depth does not.\\

\noindent
When examining the distributions for each carbonyl group individually,
the $g(r)$ profile of the terminal –COOH group (blue line) shows a
broader first hydration shell compared to the carbonyl groups of ALA1
and ALA2 while the second shell remains similar but noticeably shifted
for ML/MM simulations. This difference likely arises from its position
at the C-terminus, where the presence of a nearby hydroxyl (–OH) group
alters the local electrostatic environment, leading to distinct
hydration characteristics relative to the internal carbonyl oxygens.\\

\begin{figure}[h!]
    \centering
    \includegraphics[width=0.45\linewidth]{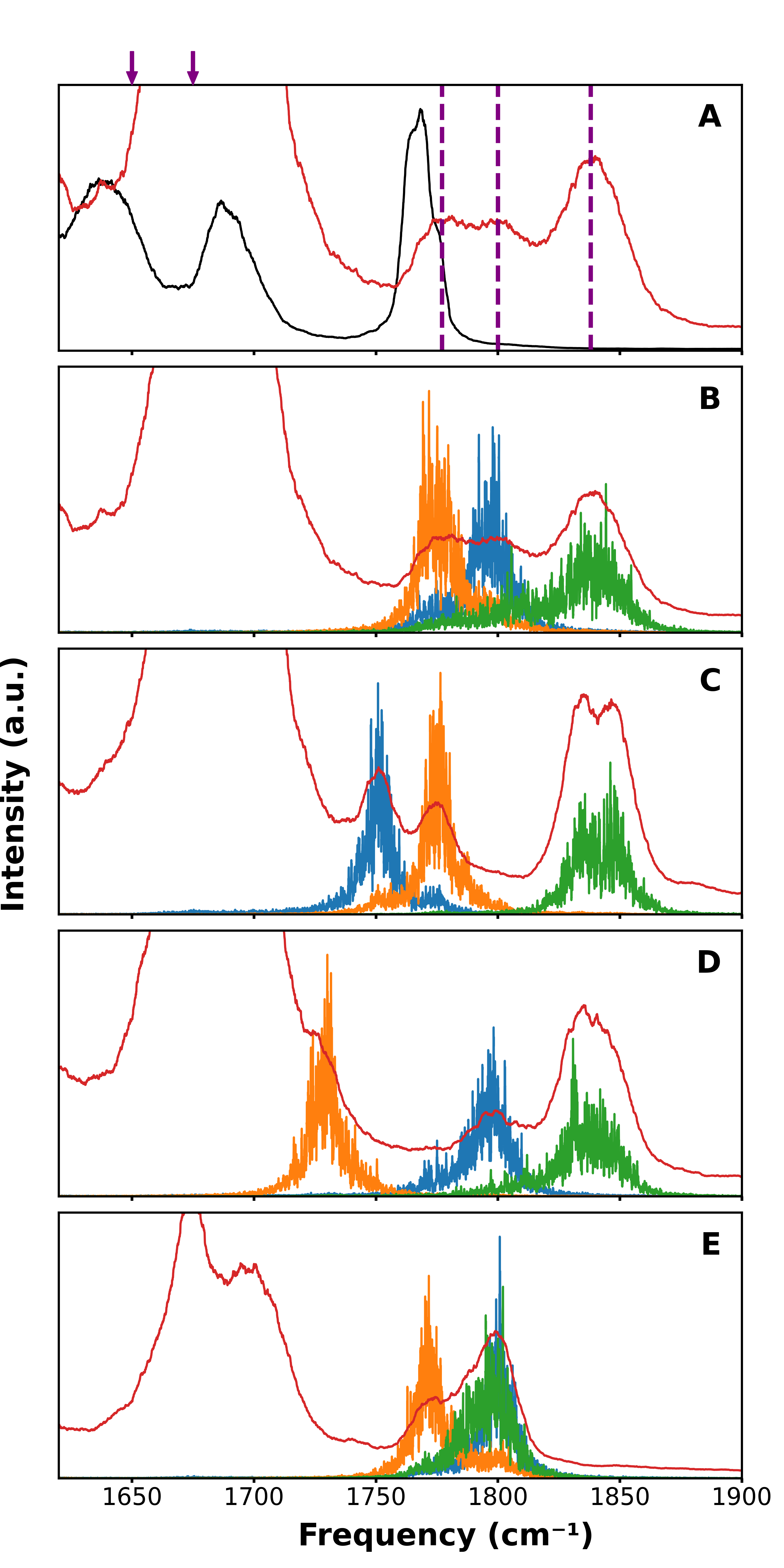}
    \caption{Comparison of computed IR and power spectra (PS) for the
      cationic AAA and measurements.\cite{woutersen:2000} Panel A: IR
      spectra from ML/MM-MD (red trace) and simulations using CGenFF
      (black trace). Panel B: IR spectrum from ML/MM-MD simulations
      together with corresponding power spectra for --C=O group in
      ALA1, ALA2, and ALA3 (blue, orange, green). Panels C to E: Power
      spectra and IR spectra for the isotopically substituted
      $^{13}$C=O at ALA1 (C), ALA2 (D), and ALA3 (E), illustrating the
      spectral shifts induced by the isotopic substitution for each
      residue in ML/MM-MD simulations. The two arrows on top of Panel
      A indicate the experimentally measured line positions at 1650
      and 1675 cm$^{-1}$, split by 25 cm$^{-1}$.\cite{woutersen:2000}
      The dashed vertical lines are shifted to best overlap with the
      doublet-structure at 1780, 1800 and 1839 cm$^{-1}$ which leads
      to a splitting of 20 cm$^{-1}$ from the ML/MM-MD
      simulations. For simulations using the CGenFF, see Figure
      \ref{sifig:aaa-cgenff}.}
    \label{fig:ir_iso}
\end{figure}

\noindent
Next, the topography of the accessible conformational landscape for
cationic AAA in the space of the two Ramachandran angles was analyzed,
see Figure \ref{fig:aaa_fes}. The aim of this analysis was to
characterize the accessible backbone conformations and to estimate the
relative free energies of different regions in $[\Phi,\Psi]$
space. For this, constraints on the $[\Phi_1,\Psi_1]-$ and
$[\Phi_2,\Psi_2]$ angles were applied separately, covering the entire
interval from $[-180,180]^\circ$ in steps of $10^\circ$. This leads to
a grid comprising 1369 grid points. For each constrained configuration
an equilibrium simulation was run for 100 ps after which the
constraint was released and the dynamics was continued for 1 ns. At
the end of these simulations the last frame of each relaxed
configuration was saved and used to build the cumulative probability
function $P(\Phi,\Psi)$ from which a pseudo-free energy surface
$\tilde{G}(\Phi,\Psi) = -\frac{1}{k_{\rm B}T} \times
\ln[P(\Phi,\Psi)]$ was estimated, see Figure \ref{fig:aaa_fes}.\\

\noindent
The wells in the FES, which represent the most stable configurations
visited on the 1 ns time scale at a given temperature, show a
distribution similar to the known low-energy regions on the
Ramachandran plot, see Figure \ref{fig:ramachandran_aaa}. This
similarity arises because both ultimately describe the energetically
preferred backbone conformations of the peptide. However, the
pseudo-FES typically resolves a larger number of distinct minima than
the Ramachandran plot, reflecting not only steric preferences but also
entropic contributions from intramolecular and solvent
interactions. Thus, while the Ramachandran plot provides a simplified,
steric-based view, the pseudo-FES can offer a more detailed
thermodynamic landscape of the conformational states first identified
by Ramachandran.\\

\noindent
Finally, the infrared spectroscopy of hydrated cationic AAA is
considered.  Figure \ref{fig:ir_iso} compares the computed IR and
power spectra (PS) for cationic AAA tripeptide with experimental
measurements.\cite{woutersen:2000} Panel A displays the IR spectra
obtained from ML/MM-MD simulations (red trace) and from simulations
using the CGenFF energy function (black trace). The ML/MM-MD spectrum
exhibits a characteristic doublet feature that corresponds closely to
the experimentally observed bands at 1650 and 1675 cm$^{-1}$ (magenta
arrows), separated by approximately 25 cm$^{-1}$. More recent
measurements\cite{Tokmakoff2018} reported these bands at 1650 and 1671
cm$^{-1}$. The measured position for the -COOH vibration was at 1725
cm$^{-1}$.\cite{woutersen:2000} The dashed vertical lines in Panel A
indicate the shifted positions of the measured doublet peaks to best
overlap with the computed amide-I peak maxima from the ML/MM-MD
simulations. These are at 1780, 1800, and 1839 cm$^{-1}$ which gives a
splitting of $\sim 20$ cm$^{-1}$ for the amide-I band. If the
frequencies obtained from the MP2 6-31G(d,p) calculations are scaled
by 0.937 (for harmonic frequencies), following established
procedures,\cite{radom:1996,cccbdb} the two amide-I bands appear at
1668 and 1687 cm$^{-1}$, in excellent agreement with the experimental
data.\\

\noindent
For unambiguous identification, Figures \ref{fig:ir_iso}C–E present
the power and IR spectra for the isotopically substituted $^{13}$C=O
groups at ALA1 (blue), ALA2 (orange), and ALA3 (green),
respectively. Isotopic substitution results in distinct red shifts in
the vibrational frequencies for each residue, demonstrating that the
ML/MM-MD simulations capture the local variations in vibrational
coupling and sensitivity to isotopic perturbation. Earlier
experimental work\cite{Tokmakoff2018} also investigated isotopically
substituted $^{13}$C=O groups at ALA1, ALA2, and ALA3, and the
reported peak positions align well with the present findings after
scaling the frequencies, see above.\\

\noindent
Overall, the present simulations for hydrated AAA using a ML/MM-PES
based on MP2/6-31G(d,p) reference calculations yield a split IR
spectrum in the amide-I region with a splitting of 25 cm$^{-1}$ which
is consistent with measurements. In addition, the conformational space
sampled by these simulations supports earlier findings that,
predominantly, cationic AAA in water adopts $\beta-$sheet and PPII
conformations. This suggests that the present approach is expected to
provide valuable information on the conformational sampling and
spectroscopic properties of tripeptides in solution. This is next
applied to a less-well characterized tripeptide: AMA in gas phase.\\

\subsection*{The AMA Tripeptide}
The second system considered in the present work is the AMA tripeptide
in the gas phase. For this, a new ML-PES was trained on
RI-MP2/[def2-SVP + def2-SVP/C] reference data using the PhysNet
architecture\cite{MM.physnet:2019} and the Asparagus
environment.\cite{MM.asparagus:2025} Molecular Dynamics simulations
were carried out in the gas phase, using the CGenFF energy
function\cite{cgenff} and the ML-PES to characterize the
conformational landscape in $[\Phi,\Psi]-$space and to obtain the
gas-phase IR spectra.\\

\begin{figure}[H]
    \centering
    \includegraphics[width=0.9\textwidth]{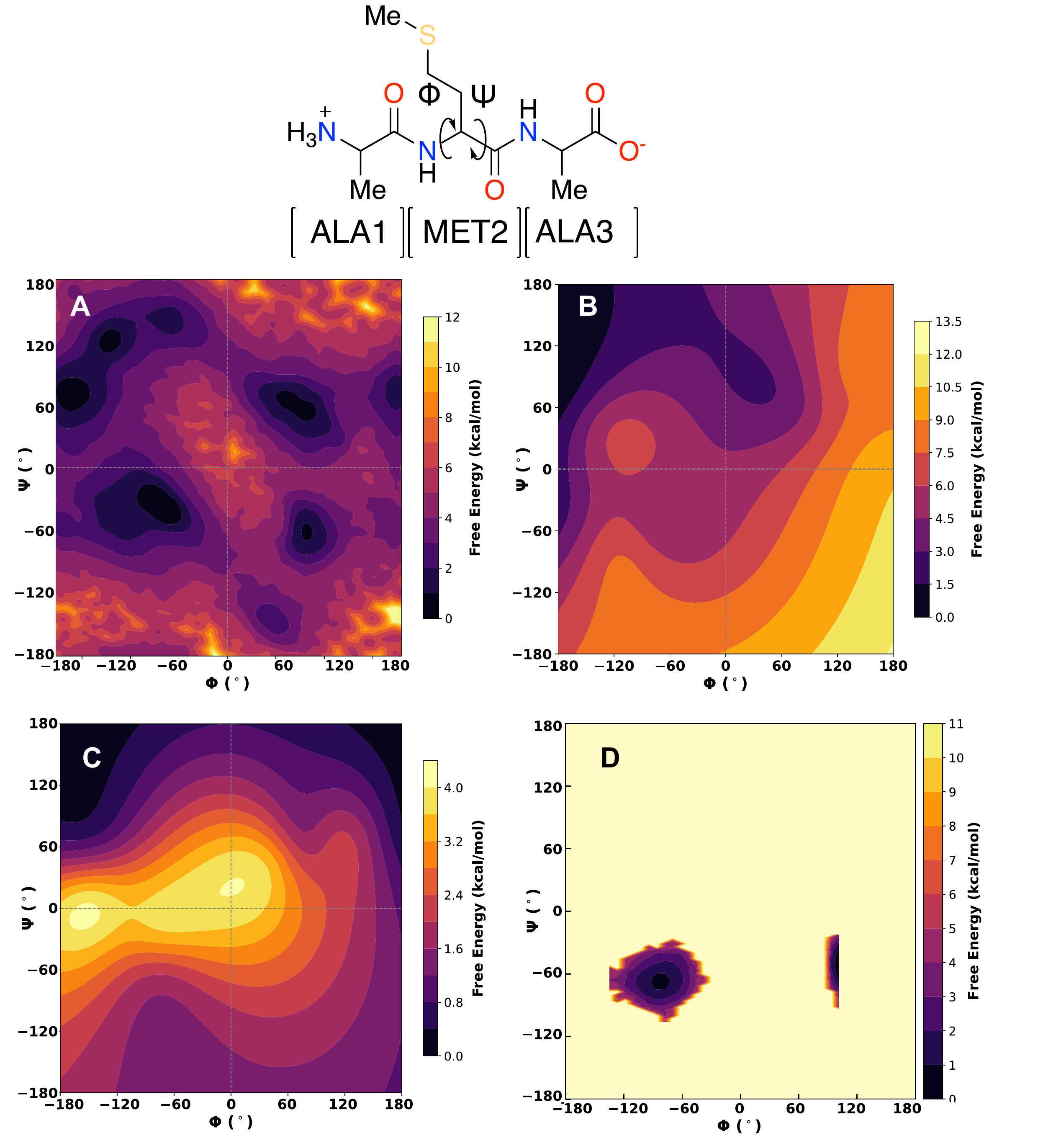}
    \caption{Top: structure of zwitterionic AMA with labelled and
      dihedral angles $[\Phi,\Psi]$ indicated.  Panel A: Dihedral
      distribution for zwitterionic AMA in the gas phase obtained from
      relaxed dynamics after releasing constraints using the CGenFF
      energy function. Darker regions indicate more populated
      conformations. Panel B: Two-dimensional free energy surface at
      300 K as a function of the $[\Phi,\Psi]$ dihedral angles
      obtained from REMD simulation for zwitterionic AMA using
      CGenFF. A single minimum near the PPII structure is found
      (black). Panel C: As for panel B but using softened XH-bond
      potentials for generating training data for the ML-PES. Panel D:
      Dihedral angle distribution from pyCHARMM MD simulations, 200 ps
      in length at 300 K, using the ML-PES for AMA in the gas
      phase. This simulations started from extended, zwitterionic AMA
      but ring-closure and neutralization already occurred during
      minimization. Hence, this landscape is for neutral AMA. Two
      wells centered at $[\Phi=-90,\Psi=-60]^\circ$ and
      $[\Phi=100,\Psi=-50]^\circ$ are found.}
\label{fig:fes-ama}
\end{figure}

\noindent
First, the conformational landscape for zwitterionic AMA (Figure
\ref{fig:fes-ama}A) sampled from the CGenFF simulations was
characterized, see Figure \ref{fig:fes-ama}A. Following the procedure
for AAA, the $[\Phi_1, \Psi_1]-$angles were constrained across the
entire interval from $[-180,180]^\circ$ in steps of $10^\circ$ and
equilibration MD simulations were run at 300 K for 100
ps. Subsequently, the constraints were released and the relaxation
dynamics was followed for 1 ns. Snapshots were written every 1 ps from
which the cumulative distribution function $P(\Phi,\Psi)$ was
generated. Inverting the Boltzmann-relationship $P(q) \sim \exp{-\beta
  G(q)}$ yields a landscape $\tilde{G}(\Phi,\Psi)$ that illustrates
the population distribution after releasing the constraints, see
Figure \ref{fig:fes-ama}A. It should be noted that
$\tilde{G}(\Phi,\Psi)$ is not an equilibrium free energy surface
$G(\Phi,\Psi)$ but rather characterizes the system on the 1 ns time
scale and at 300 K after releasing the constraints. On the other hand,
such a procedure provides a meaningful first overview of possible
minimum energy structures. Six distinct low-energy basins (black
densities), corresponding to different conformations of zwitterionic
AMA are found from this approach. The include PPII, $\beta-$sheet,
right- and left-handed $\alpha-$helical and an unlabelled structure
centered at $[\Phi = 90, \Psi = -60]^\circ$.\\

\noindent
Next, the machine-learned energy function was constructed. First, REMD
simulations for zwitterionic and neutral AMA using CGenFF were run
with replicas at $T \in [300, 350, 400, 450, \\ 500, 550, 600, 650]$
K. In addition, REMD at the same temperatures for zwitterionic AMA
with softened XH-bond stretching potentials was carried out. It has
been found that sampling the XH-bonds (X = C, N, O) sufficiently
broadly is important for a stable ML-PES using PhysNet. Therefore, a
soft Morse oscillator was used for all bonds involving hydrogen
atoms. Using MBAR\cite{shirts:2008} the FES $G(\Phi,\Psi)$ was
constructed from the aggregate of the sampled structures during REMD,
see Figure \ref{fig:fes-ama}C. Panels B and C report results from REMD
simulations for zwitterionic AMA without applying Morse oscillators
(Panel B) and with applying them (Panel C). There are similarities in
terms of regions covered during REMD such as the area $[\Phi \sim
  -180, \Psi \sim 180]^\circ$, typical for a $\beta$-sheet
conformation. On the other hand, $G(\Phi,\Psi)$ from using softened
X-H bonds is considerably flatter (maximum $G(\Phi,\Psi) \sim 4$
kcal/mol) than that using the conventional CGenFF energy function
(maximum $G(\Phi,\Psi) \sim 12$ kcal/mol). From the REMD simulations
with softened XH-bonds 20000 structures were extracted for training
the ML-PES.\\

\noindent
Using a 80/10/10 \% split, the data set from the REMD simulations was
used together with the Asparagus suite\cite{MM.asparagus:2025} to
train the ML-PES. The quality of the final model is characterized by a
mean average error for energies and forces of 0.27 kcal/mol and 0.41
(kcal/mol)$\cdot$ \AA$^{-1}$; the corresponding root mean squared
errors are 0.38 kcal/mol and 0.63 (kcal/mol)$\cdot$ \AA$^{-1}$,
respectively. The overall performance for energies is reported in
Figure \ref{sifig:ama-nn}. Because free dynamics of zwitterionic AMA
using the ML-PES resulted in ring closure and subsequent H-transfer to
form neutral AMA, the following simulations describe the dynamics of
the neutral species. It is important to note that the NN-PES was
trained on both, zwitterionic and neutral forms of AMA. This allowed
stable MD simulations in the gas phase for both tautomers. The
stability of this ML-PES was further assessed from diffusion Monte
Carlo (DMC) simulations using a step size of 0.1 \AA\/ and
accumulating $2.5 \cdot10^6$ structures. No holes, characterized by
the fact that the energy of a particular sample is below the energy of
the global minimum, was found. The minimum energy structure adopted by
neutral AMA using the ML-PES is characterized by
$[\Phi=-87,\Psi=-72]^\circ$.\\

\noindent
Using the trained PES, ML/MD simulations 200 ps in length were carried
out in the gas phase. The initial structure before heating was that of
zwitter-ionic AMA which neutralized already during minimization. The
dihedral angle distribution $P(\Phi,\Psi)$ reported in Figure
\ref{fig:fes-ama}D revealed primarily the presence of a
$\alpha-$helical structure (well 1). However, constraining $[\Phi_1
  \sim 100^\circ,\Psi_1 \sim 100^\circ]$, a second minimum (well 2)
emerges. As zwitter-ionic AMA spontaneously converts to neutral AMA
from simulations using the ML-PES in the gas phase, it is also of
interest to consider the free energy surface for neutral AMA from
simulations using CGenFF, see Figure
\ref{sifig:FES_neutral}. Neutralizing the two ends reshapes
$G(\Phi,\Psi)$ such that PPII and $\beta-$sheet structures are
favourable with low-energy conformations extending into the region of
$\alpha-$helical structures.\\

\begin{figure}[h!]
    \centering
    \includegraphics[width=0.6\textwidth]{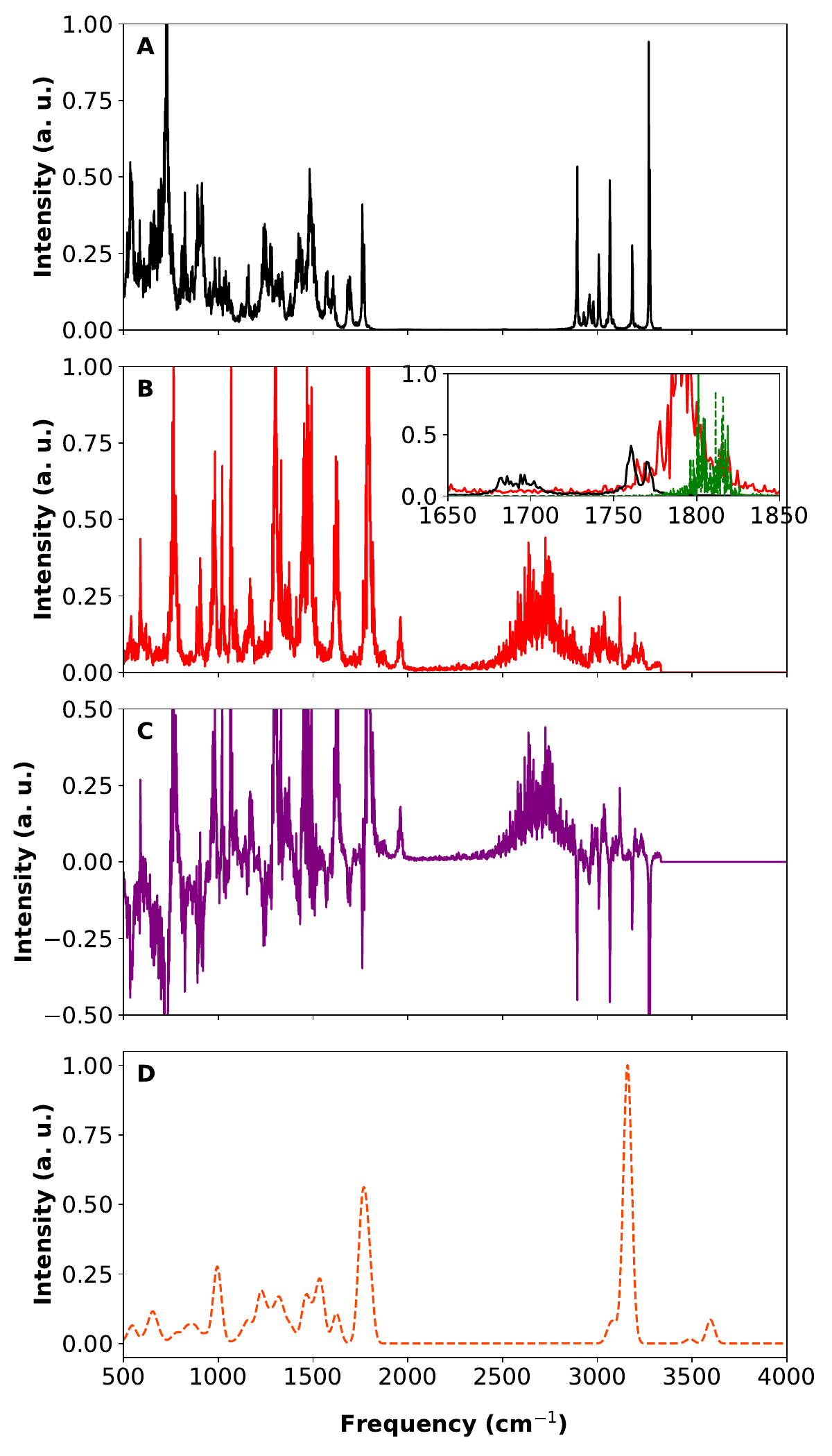}
    \caption{Infrared spectra for neutral AMA. Panel A: Simulations
      using CGenFF. Panel B: Simulation using the ML-PES sampling the
      minimum around $[\Phi=-90,\Psi=-60]^\circ$ in Figure
      \ref{fig:fes-ama}D. Panel C: difference spectrum between panels
      A and B. Panel D: normal mode calculation for the optimized
      structure with $[\Phi=-90,\Psi=-60]^\circ$ using the
      RI-MP2/[cc-pVTZ+cc-pVTZ/C] level of theory. The inset in panel B
      focuses on the amide-I band. Color code: black: IR-spectrum from
      CGenFF simulations; red: IR-spectrum from ML-PES simulations;
      green: CO-power spectrum for ALA1 from the ML-PES
      simulations. For a detailed view of the power spectrum, see
      Figure \ref{sifig:AMA_power_MLFF}.}
    \label{fig:ir-ama}
\end{figure}

\noindent
Figure \ref{fig:ir-ama} reports the IR spectra for neutral AMA
obtained from using CGenFF and the ML-PES as the energy
function. First, the averaged (10 independent simulations) IR spectrum
from 1 ns simulations using CGenFF is considered, see Figure
\ref{fig:ir-ama}A. As usual, the vibrations for bonds involving
H-atoms appear around 3000 cm$^{-1}$ whereas the framework modes are
below 2000 cm$^{-1}$. Of particular relevance are the amide-I
vibrations (see also AAA above) which appear at $\sim 1700$ cm$^{-1}$
as assigned from the power spectra reported in Figure
\ref{sifig:AMA_power}. It should be noted that all bonds are described
as harmonic oscillators and the IR-intensities are determined from
dipole moment autocorrelation function using the static point charges
of CGenFF.\cite{cgenff}\\

\noindent
The IR spectrum from 200 ps simulations using the ML-PES for neutral
AMA is reported in Figure \ref{fig:ir-ama}B. In Panel B, the inset
highlights the Amide-I region, comparing the ML-PES (red trace),
CGenFF (black trace), and the power spectrum of the C=O bond distance
in the ALA1 residue (green trace). A notable difference between panels
A and B is the broad feature below 3000 cm$^{-1}$ which is due to the
H-bonding interactions between the -NH$_2$ and -COOH termini in
neutral AMA, see also Figure \ref{sifig:ama-termini-power}. The
pronounced red shift by up to 500 cm$^{-1}$ for the NH- and OH-stretch
vibrations is consistent with OHO-motifs as the occur and have been,
e.g. experimentally and computationally characterized in protonated
oxalate.\cite{wolke:2015,MM.oxalate:2017,MM.oxalate:2025} Figures
\ref{fig:ir-ama}C/D report the difference spectrum (panel B - panel A)
and the normal mode spectrum neutral AMA from calculations at the
RI-MP2/[cc-pVTZ + cc-pVTZ/C] level of theory to provide some
guidance. \\

\noindent
For the second minimum with $[\Phi=100,\Psi=-50]^\circ$ in Figure
\ref{fig:fes-ama}D the IR spectrum was determined as well, see Figure
\ref{sifig:IR_AMA_MLFF}B. There are distinct differences compared with
the IR spectrum for the primary minimum at
$[\Phi=-90,\Psi=-60]^\circ$. First, the peak positions and intensities
for the modes at and below 1000 cm$^{-1}$ differ in a distinct
manner. Secondly, the amide-I region features a different number of
peaks with modified intensity distributions. And finally, the
high-frequency hydrogen-stretch region for the main minimum has a
double-peak structure below 3000 cm$^{-1}$ followed by a diffuse band
centered at 3000 cm$^{-1}$ whereas for the secondary minimum there is
a single broad absorption below 3000 cm$^{-1}$ and sharp peaks above
3000 cm$^{-1}$. Such examples for isomer-specific IR spectra indicate
how spectroscopy can be used for structure-identification.\\

\section*{Conclusion}
The present work considered the use of ML-based energy functions,
trained on MP2 reference data for characterizing the conformational
landscape and infrared spectroscopy of tripeptides. For cationic AAA
excellent agreement for the IR-spectroscopy was found - modulo an
explainable overall shift of the bands due to the level of quantum
chemical theory used. The splitting between the two bands is in almost
quantitative agreement with experiments and the ordering of the
amide-I and terminal -CO band is also consistent with experiments.\\

\noindent
For AMA the focus was on the gas phase dynamics. This was also
motivated by the finding that starting from the zwitterionic species
neutralization occurs during ML-MD dynamics in the gas phase which
necessitated the retraining of the entire ML-PES to include both
tautomers. A general observation in conceiving ML-PESs for new systems
is the fact that the inherent reactivity of such models makes it
difficult to decide {\it a priori} what the properties of a suitable
training set are. For AMA the dominant structure in the gas phase from
ML-MD simulations is $\alpha-$helical. A second minimum was found
which, however, was never reached from equilibrium simulations unless
the dynamics was initiated very close to
$[\Phi=100,\Psi=-50]^\circ$. For AMA the ML-PES is now suitable for
characterizing the conformational landscape and spectroscopy in
solution.\\

\noindent
In conclusion, the present work demonstrates that stable and
meaningful ML-MD and ML/MM-MD simulations at the MP2-level of theory
are possible on the multi-nanosecond time scale in the gas phase and
in solution. This is extensible to general tripeptides XYZ.\\

\section*{Author Contributions}

\section*{Data Availability} 
The reference data that allow to reproduce the findings of this study
are openly available at \url{https://github.com/MMunibas/aaa.ama}.\\

\section*{Acknowledgment}
Financial support from the Swiss National Science Foundation through
grants $200020\_219779$ (MM), $200021\_215088$ (MM), the University of
Basel (MM) is gratefully acknowledged.

\bibliography{ala3.tidy}

\clearpage

\renewcommand{\thepage}{S\arabic{page}}
\renewcommand{\thetable}{S\arabic{table}}
\renewcommand{\thefigure}{S\arabic{figure}}
\renewcommand{\theequation}{S\arabic{equation}}
\renewcommand{\thesection}{S\arabic{section}} 
\setcounter{figure}{0}  
\setcounter{section}{0}  
\setcounter{table}{0}

\section*{Supporting Material}

\section{AAA}

\begin{figure}[h!]
    \centering
    \includegraphics[width=0.7\textwidth]{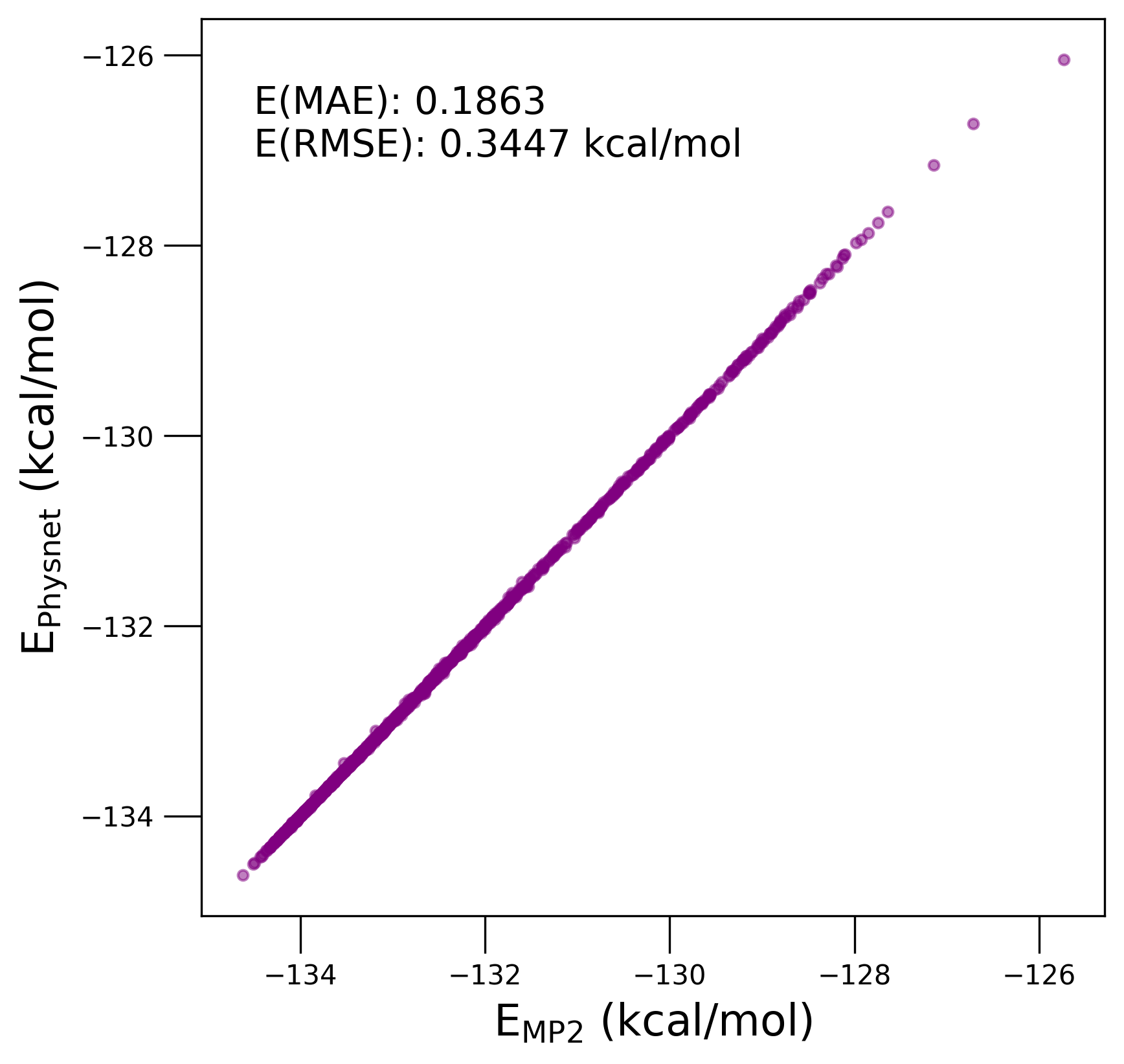}
    \caption{Performance of the PhysNet ML-PES on reference data
      determined at the MP2/6-31G(d,p) level of theory.}
    \label{sifig:aaa-nn}
\end{figure}

\begin{figure}
    \centering
    \includegraphics[width=0.5\linewidth]{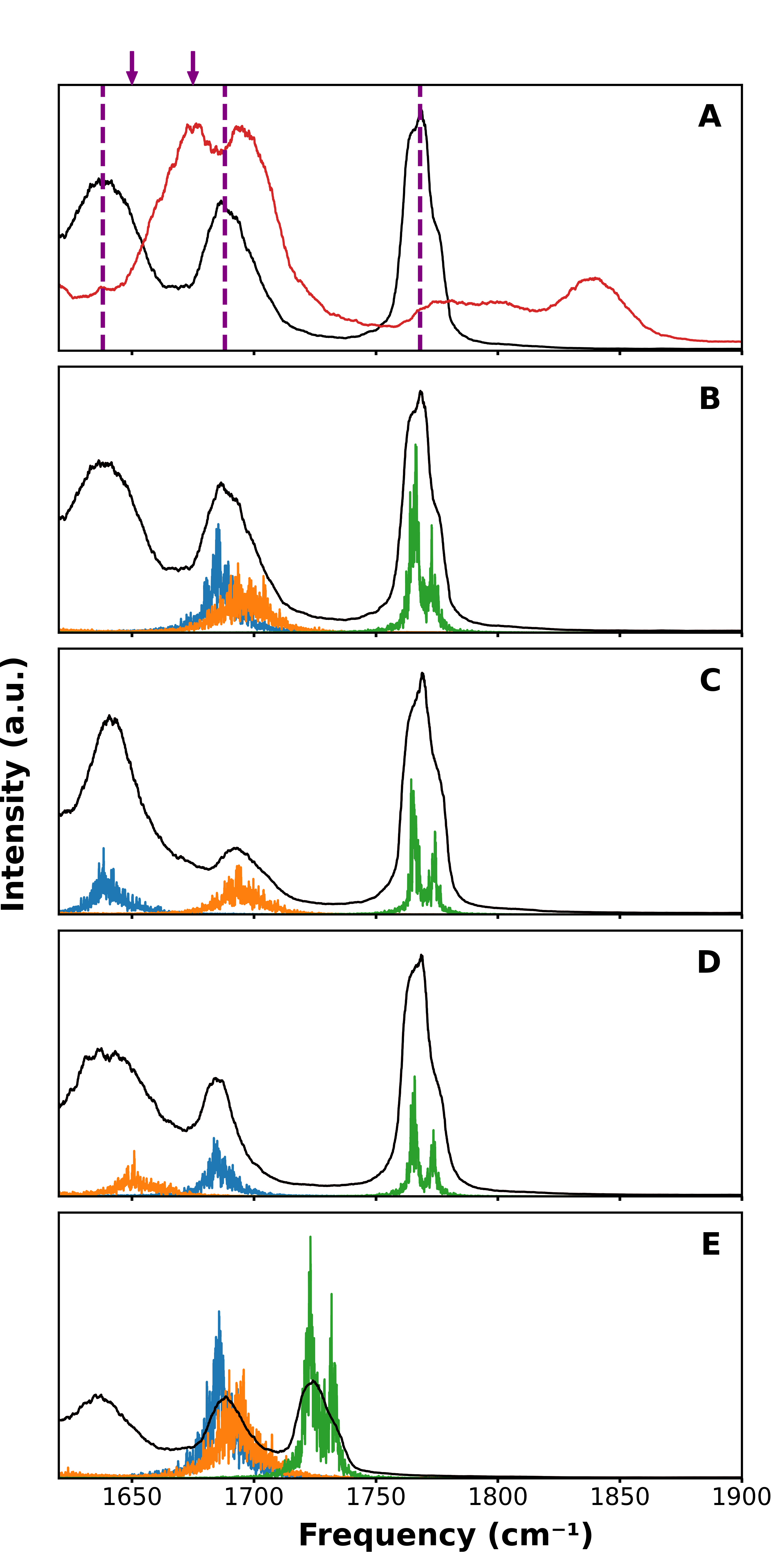}
    \caption{Comparison of IR and power spectra (PS) for cationic
      AAA. Panel A: IR spectra from simulations using CGenFF (black)
      and the ML-PES (red). Panel B: Black spectrum from panel A
      together with corresponding power spectra for the three -CO
      groups, see Figure \ref{fig:ir_iso}. Panels C to E: Power
      spectra and IR spectra for isotopically substituted $^{13}$C=O
      at ALA1 (C), ALA2 (D), and ALA3 (E), illustrating the spectral
      shifts induced by the isotopic substitution for each residue.}
    \label{sifig:aaa-cgenff}
\end{figure}

\newpage

\section{AMA}
\begin{figure}[h!]
    \centering
    \includegraphics[width=1.0\textwidth]{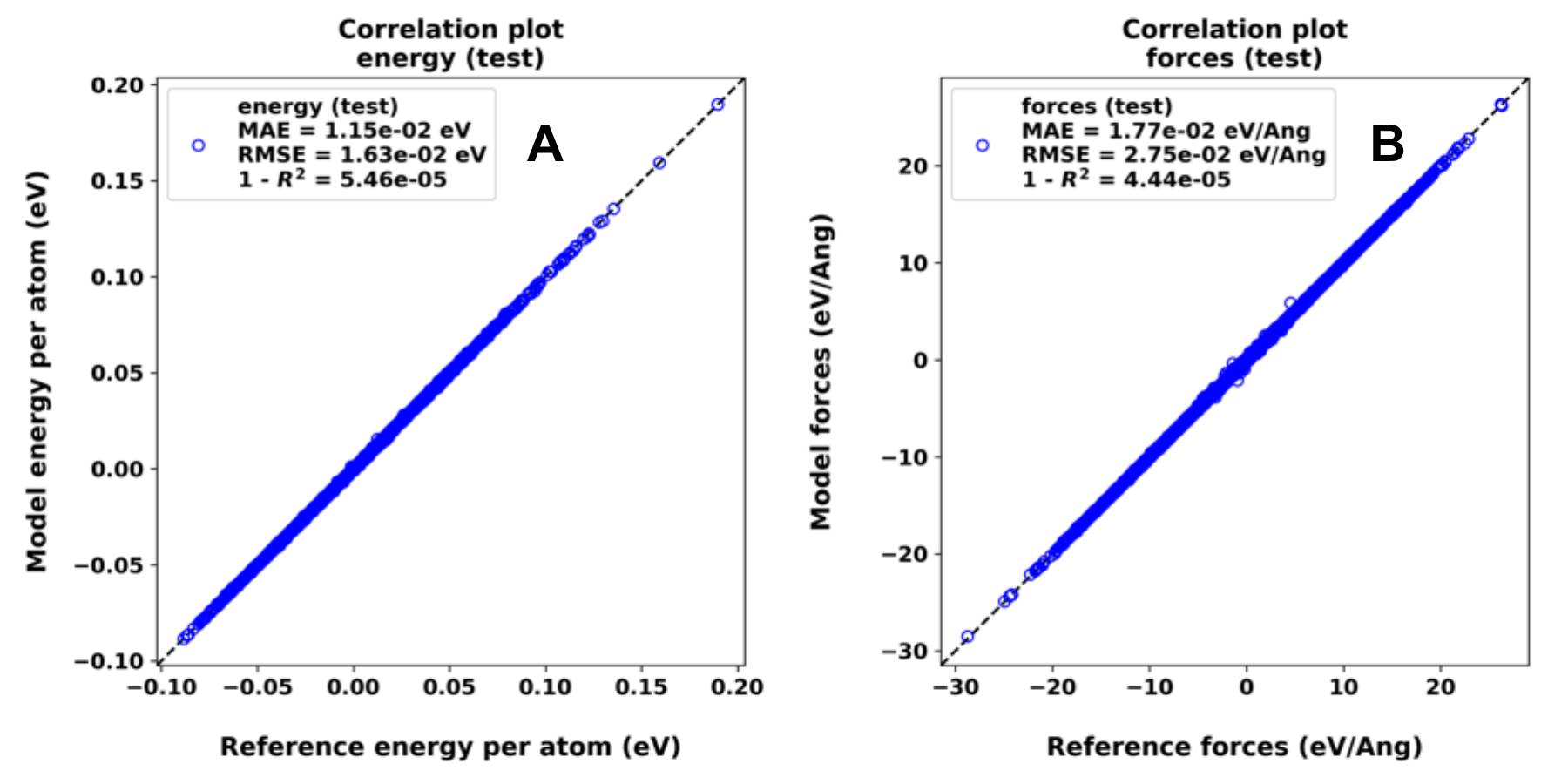}
    \caption{Correlation plot for energies (Panel A) and forces (Panel
      B) on the test set for the AMA model. The data set contains
      both, zwitterionic and neutral AMA, and the reference
      calculations were carried out at the RI-MP2/[def2-SVP +
        def2-SVP/C] level of theory.}
    \label{sifig:ama-nn}
\end{figure}

\begin{figure}[h!]
    \centering
    \includegraphics[width=1.0\textwidth]{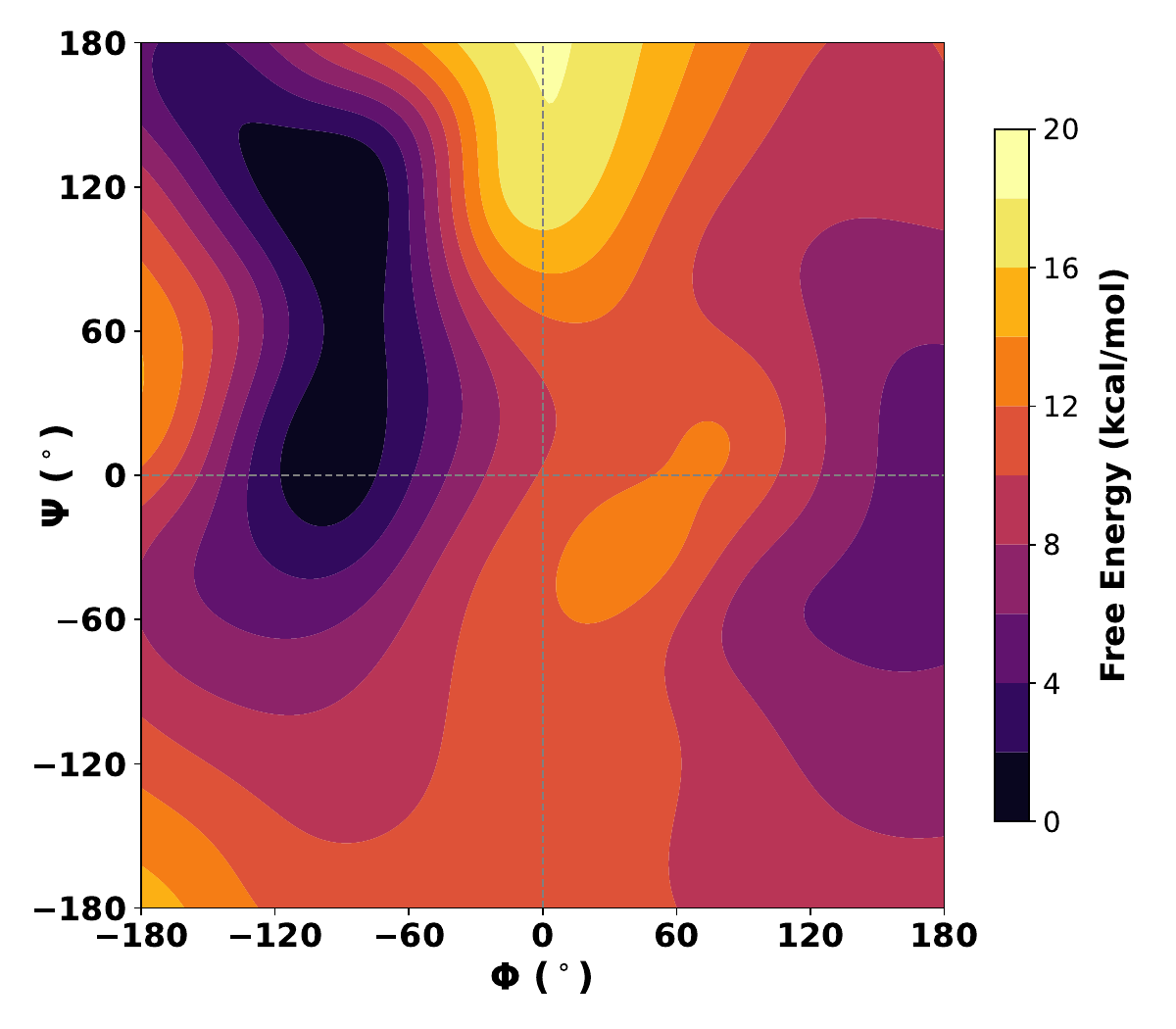}
    \caption{The free energy surface for neutral AMA from REMD
      simulations using the CGenFF energy function. Compare this with
      Figure \ref{fig:fes-ama}B for the zwitterion.}
    \label{sifig:FES_neutral}
\end{figure}

\begin{figure}[h!]
    \centering
    \includegraphics[width=0.7\textwidth]{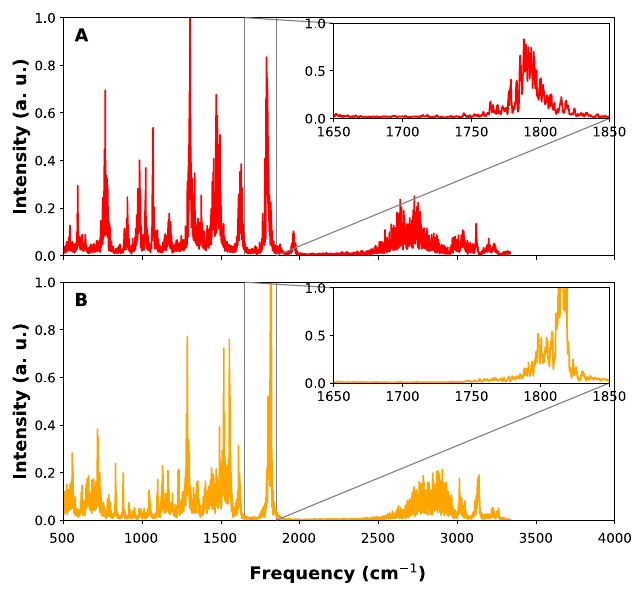}
    \caption{Computed IR spectra from ML-MD simulations using the
      NN-PES. Panel A: Spectrum from sampling the primary minimum with
      $[\Phi = -90, \Psi = -60]^\circ$. Panel B: Spectrum from
      sampling the minimum at $[\Phi = 100, \Psi = -50]^\circ$, see
      Figure \ref{fig:fes-ama}D. The inset shows an enlargement of the
      amide-I region.}
    \label{sifig:IR_AMA_MLFF}
\end{figure}

\begin{figure}[h!]
    \centering
    \includegraphics[width=1.0\textwidth]{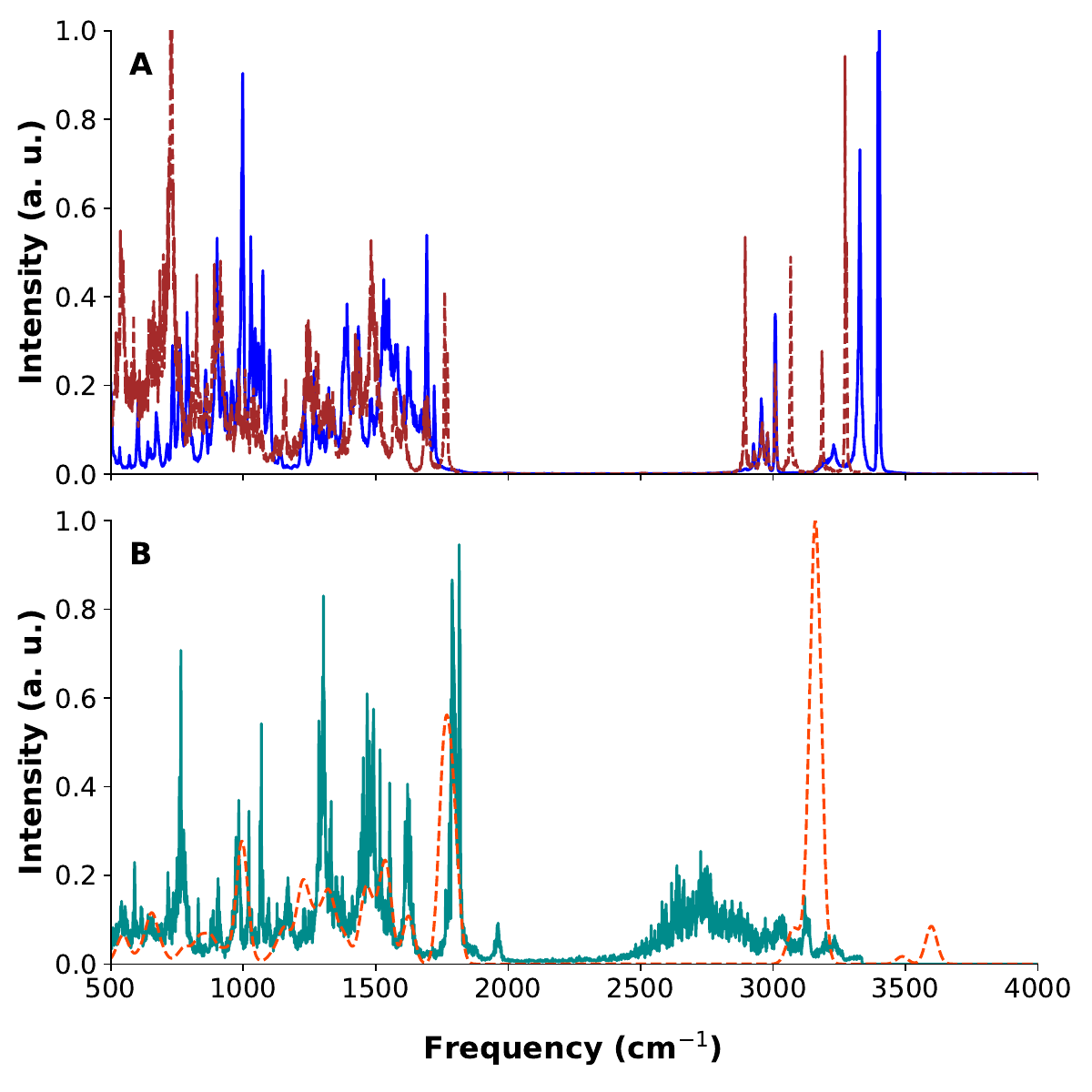}
    \caption{Comparison between several computed IR spectra for
      AMA. Panel A: IR spectrum from CGenFF MD simulations for
      zwitterionic AMA (blue trace) and for neutral AMA (red trace);
      Panel B: IR spectrum from pyCHARMM simulations for neutral and
      cyclic AMA using the trained ML-PES (cyan trace) together with
      the normal mode spectrum from QM calculations at the
      RI-MP2/[cc-pVTZ+cc-pVTZ/C] level of theory (orange dashed
      line). Due to the H-bonds between the --NH$_2$ and --COOH
      termini and the pronounced anharmonicities of the NH- and
      OH-bonds, the IR spectra from the MD simulations are strongly
      red shifted. The power spectra in Figure
      \ref{sifig:ama-termini-power} confirm these assignments.}
    \label{sifig:IR_AMA_overlap}
\end{figure}

\begin{figure}[h!]
    \centering
    \includegraphics[width=1.0\textwidth]{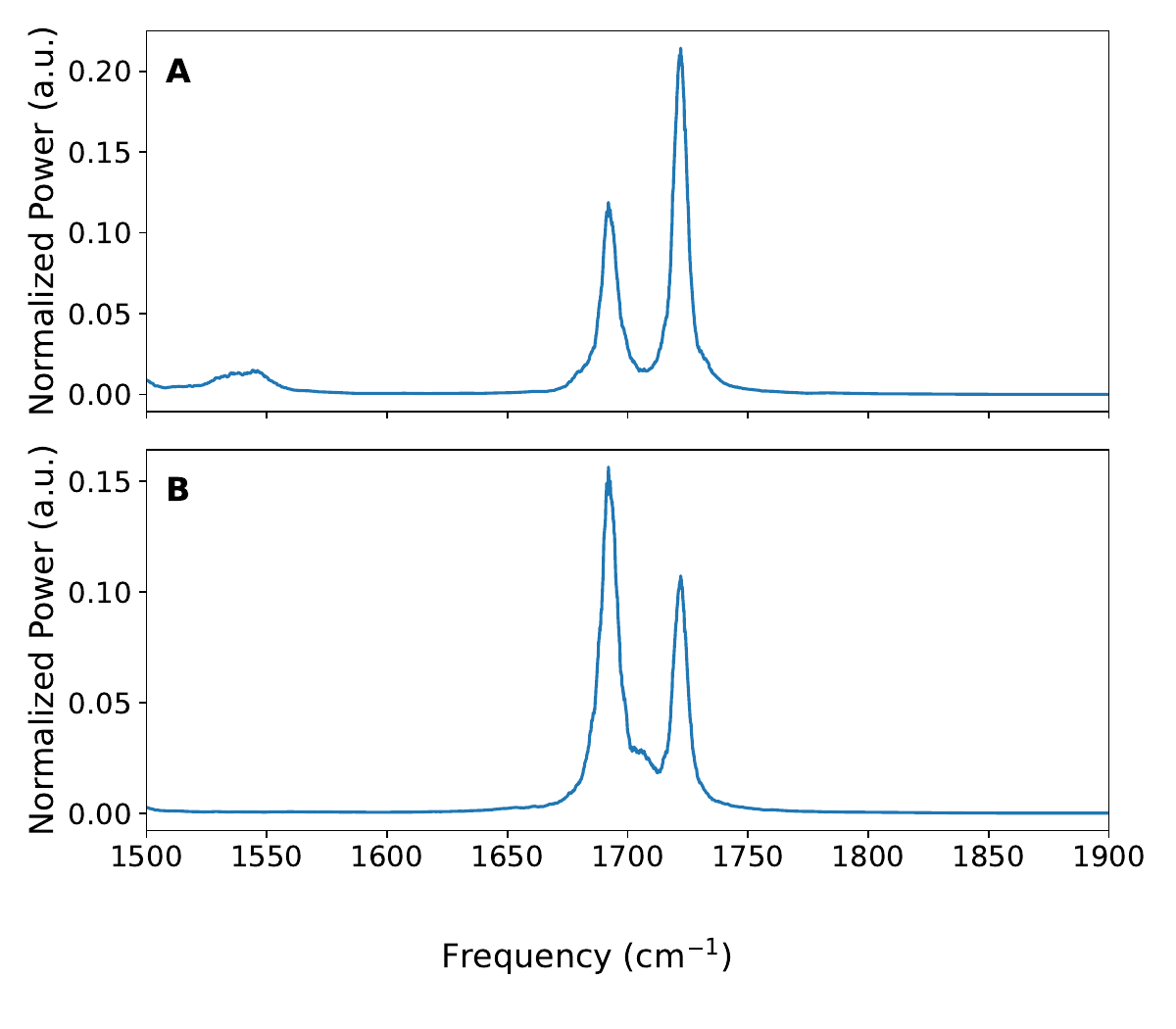}
    \caption{Power spectra based on C=O distances in ALA1 and MET2
      from simulations using the CGenFF energy function for AMA.}
    \label{sifig:AMA_power}
\end{figure}

\begin{figure}[h!]
    \centering
    \includegraphics[width=1.0\textwidth]{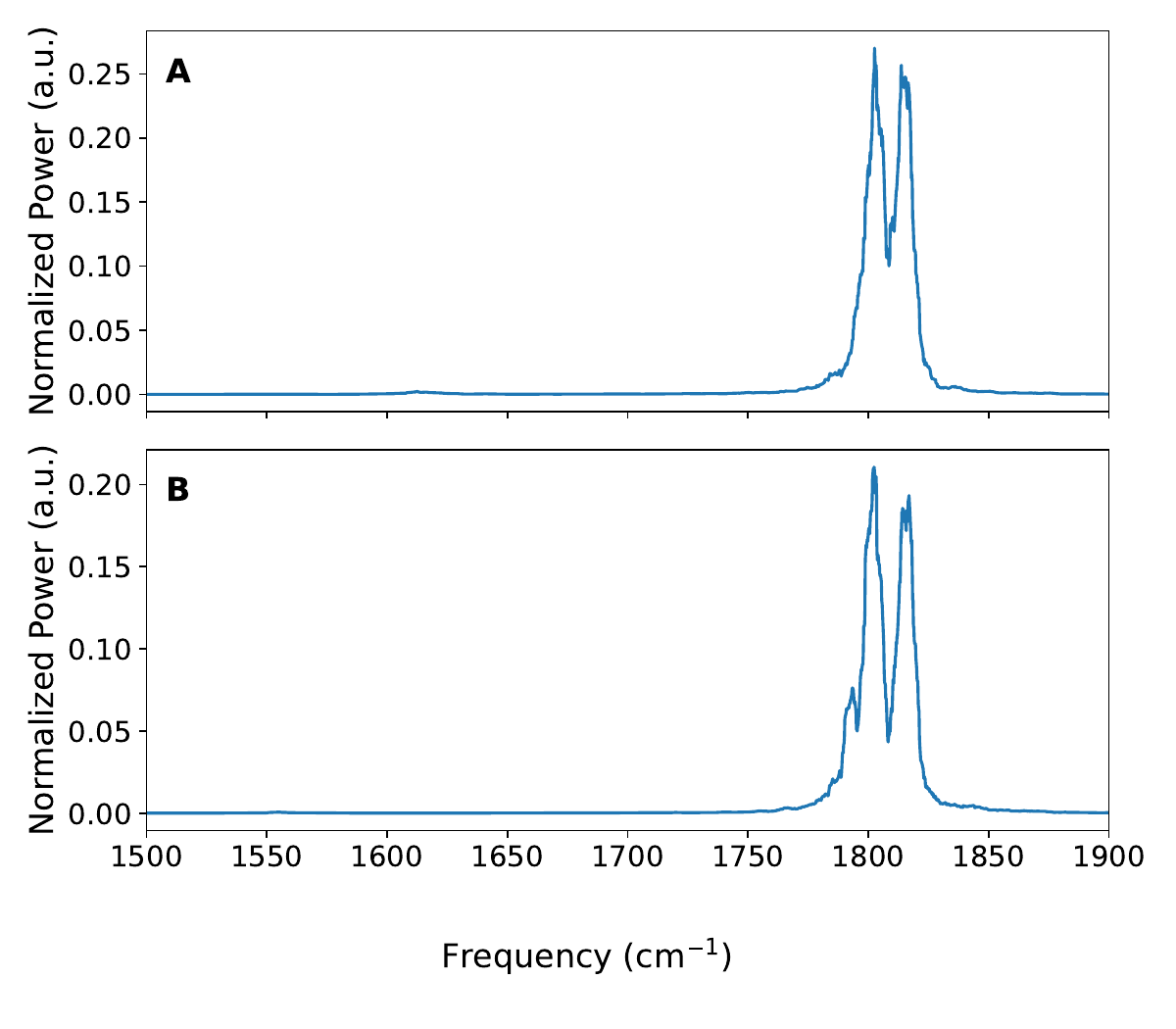}
    \caption{Power spectra based on C=O distances in ALA1 and MET2
      from simulations using the ML-PES for AMA. As was already found
      for AAA, the frequencies are shifted to the blue due to using
      the RI-MP2 method, see Figure \ref{fig:ir_iso}.}
    \label{sifig:AMA_power_MLFF}
\end{figure}

\begin{figure}[h!]
    \centering
    \includegraphics[width=1.0\textwidth]{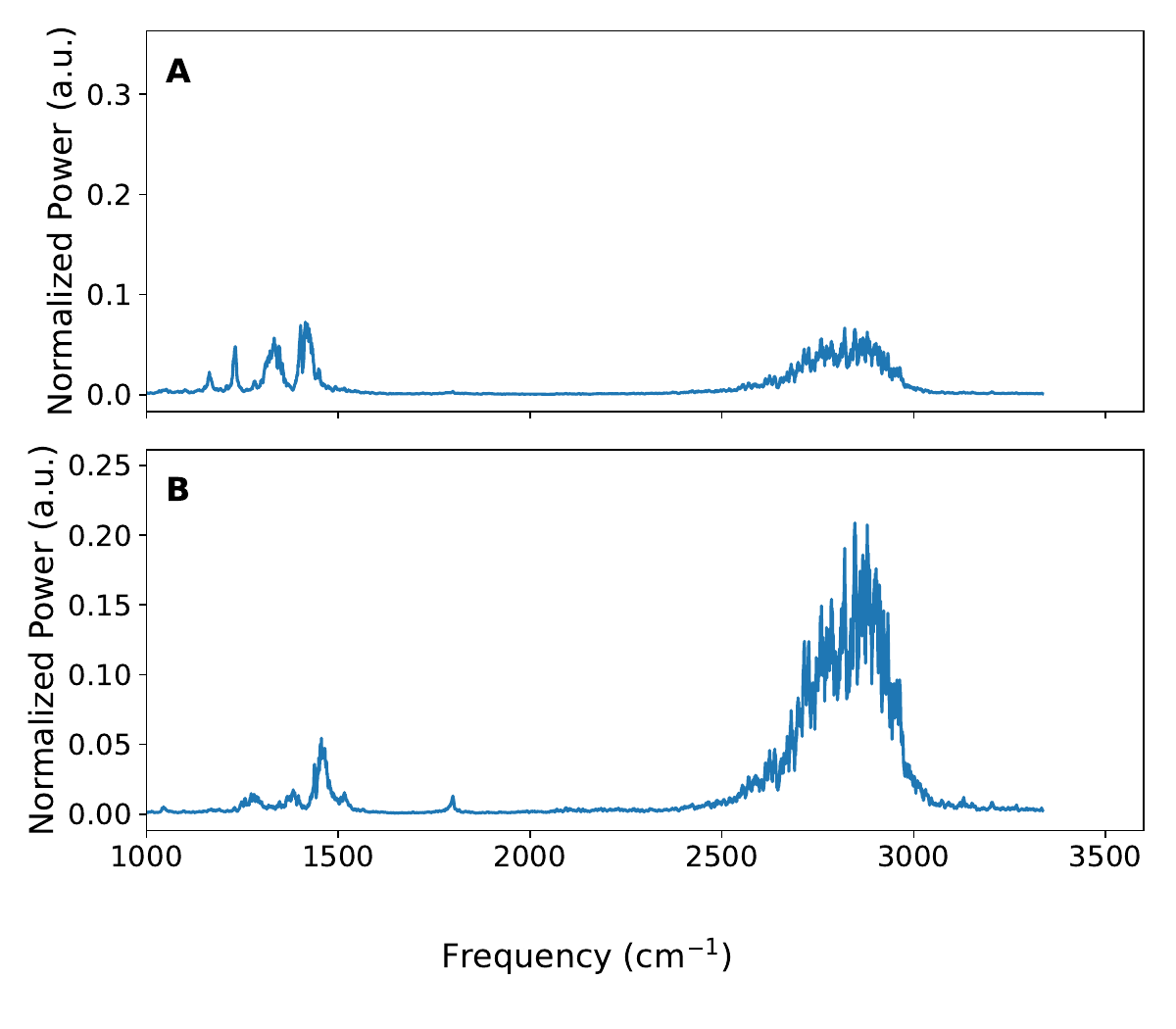}
    \caption{Power spectra for NH-stretch (Panel A) and for OH-stretch
      (Panel B) coordinates from simulations using the ML-PES. The
      prominent red shift compared to usual frequency ranges is due to
      hydrogen bonding following ring closure in neutral AMA. See also
      IR spectrum in Figures \ref{fig:ir-ama} and
      \ref{sifig:IR_AMA_MLFF}.}
    \label{sifig:ama-termini-power}
\end{figure}

\end{document}